\documentclass[12pt,letterpaper]{article}
\usepackage{epsfig,setspace,amsmath,epsf,amssymb,bm,theorem,cite, graphicx, epstopdf, algorithm, algpseudocode,float,color}
\usepackage{multirow}
\usepackage[table,xcdraw]{xcolor}

\newtheorem{theorem}{Theorem}

\newtheorem{remark}{Remark}
\newtheorem{lemma}{Lemma}

\newenvironment{Proof}[1]{\medskip\par\noindent{\bf Proof:\,}\,#1}{{\mbox{\,$\blacksquare$}\par}}

\allowdisplaybreaks

\begin{document}
	
\title{The Capacity of Private Information Retrieval from Decentralized Uncoded Caching Databases\thanks{This work was supported by NSF Grants CNS 15-26608, CCF 17-13977 and ECCS 18-07348.}}
	
\author{Yi-Peng Wei \quad Batuhan Arasli \quad Karim Banawan \quad Sennur Ulukus\\
	\normalsize Department of Electrical and Computer Engineering\\
	\normalsize University of Maryland, College Park, MD 20742 \\
	\normalsize {\it ypwei@umd.edu ~~ barasli@umd.edu ~~ kbanawan@umd.edu ~~ ulukus@umd.edu} }
	
\maketitle

\begin{abstract}
We consider the private information retrieval (PIR) problem from decentralized uncoded caching databases. There are two phases in our problem setting, a caching phase, and a retrieval phase. In the caching phase, a data center containing all the $K$ files, where each file is of size $L$ bits, and several databases with storage size constraint $\mu K L$ bits exist in the system. Each database independently chooses $\mu K L$  bits out of the total $KL$ bits from the data center to cache through the same probability distribution in a decentralized manner. In the retrieval phase, a user (retriever) accesses $N$ databases in addition to the data center, and wishes to retrieve a desired file privately. We characterize the optimal normalized download cost to be $\frac{D}{L} =  \sum_{n=1}^{N+1}  \binom{N}{n-1} \mu^{n-1} (1-\mu)^{N+1-n} \left( 1+ \frac{1}{n} + \dots+ \frac{1}{n^{K-1}}  \right)$. We show that uniform and random caching scheme which is originally proposed for decentralized coded caching by Maddah-Ali and Niesen, along with Sun and Jafar retrieval scheme which is originally proposed for PIR from replicated databases surprisingly result in the lowest normalized download cost. This is the decentralized counterpart of the recent result of Attia, Kumar and Tandon for the centralized case. The converse proof contains several ingredients such as interference lower bound, induction lemma, replacing queries and answering string random variables with the content of distributed databases, the nature of decentralized uncoded caching databases, and bit marginalization of joint caching distributions.
\end{abstract}

\section{Introduction}

\begin{figure}[t]
	\centering
	\epsfig{file=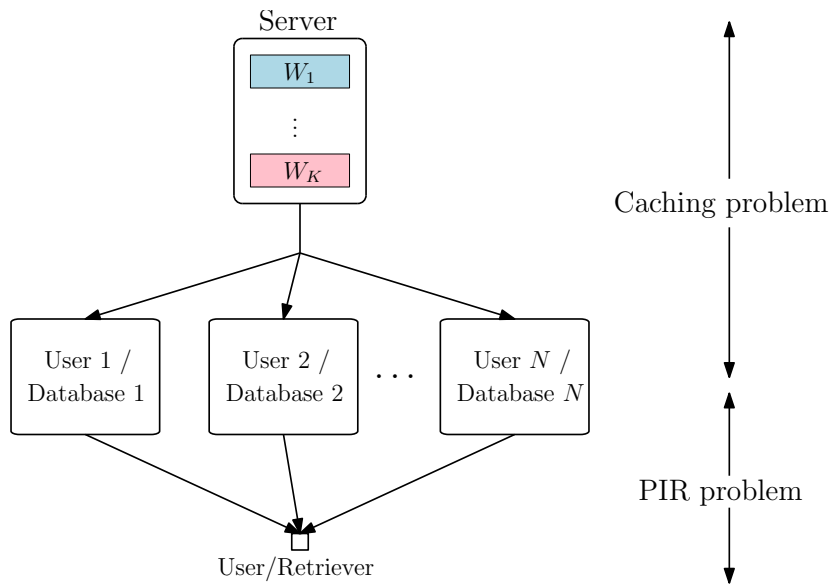,width=0.66\textwidth}
	\caption{Joint centralized caching and PIR problem.}
	\label{caching_PIR}
\end{figure}

Private information retrieval (PIR) refers to the problem of downloading a desired file from distributed databases while keeping the identity of the desired file private against the databases. In the classical setting of PIR (see Fig.~\ref{caching_PIR}), there are $N$ non-communicating databases, each storing the same set of $K$ files. The user wishes to download one of these $K$ files without letting the databases know the identity of the desired file. A simple but highly inefficient way is to download all the files from a particular database, which results in the normalized download cost of $\frac{D}{L}=K$, where $L$ is the file size and $D$ is the total number of downloaded bits from the $N$ databases. The PIR problem has originated in the computer science community \cite{ChorPIR, PIRsurvey2004, cachin1999computationally, ostrovsky2007survey, yekhanin2010private} and has drawn attention in the information theory society with early examples \cite{RamchandranPIR, unsynchonizedPIR, YamamotoPIR, VardyConf2015, RazanPIR, JafarConf2016}. Recently, Sun and Jafar \cite{JafarPIR} have characterized the optimal normalized download cost for the classical PIR problem to be $\frac{D}{L}=\left( 1+\frac{1}{N}+\dots+\frac{1}{N^{K-1}}\right)$. After \cite{JafarPIR}, many interesting variants of the classical PIR problem have been investigated in \cite{JafarColluding, symmetricPIR, KarimCoded, arbmsgPIR, codedsymmetric, MultiroundPIR, codedcolluded, codedcolludedJafar, arbitraryCollusion, MPIRjournal, codedcolludingZhang, MPIRcodedcolludingZhang, BPIRjournal, tandon2017capacity, symmetricByzantine, tajeddine2017robust, wang2017linear, kadhe2017private, wei2017fundamental, chen2017capacity, wei2017capacity, wang2017secure, sun2017_computation, Kimcache2017, mirmohseni2017private, abdul2017private, wei2017jsac, banawan2018asymmetry, chen2018asymptotic, banawan2018private, wang2018secure, attia2018capacity, wei2018capacity, tajeddine2018privatex, banawan2018noisy, jia2018cross, tian2018capacity, kumar2018private, bitar2018staircase, li2018converse, d2018one, tajeddine2018private}. Most of these previous works consider the case where the contents of the databases are fixed a priori in an uncontrollable manner, and a vast majority of them consider the case of replicated databases where each database stores the same set of $K$ files.

% JafarColluding, May 2, 2016
% symmetricPIR, Jun. 28, 2016
% KarimCoded, Sep. 26, 2016
% arbmsgPIR, Oct. 10, 2016
% codedsymmetric, Oct. 14, 2016
% MultiroundPIR, Nov. 7, 2016
% codedcolluded, Nov. 7, 2016
% codedcolludedJafar, Jan. 26, 2017
% arbitraryCollusion, Jan. 26, 2017
% Mpirjournal, Feb. 6, 2017
% codedcolludingZhang, Apr. 22, 2017
% MpircodedcolludingZhang, May. 9, 2017
% Bpirjournal, Jun. 5, 2017
% tandon2017capacity, Jun. 21, 2017   !!!
% symmetricByzantine, Jul. 7, 2017
% tajeddine2017robust, Jul. 31 2017
% wang2017linear, Aug. 17, 2017
% kadhe2017private, Sep. 1, 2017 ~~~
% wei2017fundamental, Sep. 4, 2017 ~~~
% chen2017capacity, Spe. 10, 2017 ~~~
% wei2017capacity, Oct. 2, 2017 ~~~
% wang2017secure, Oct. 3, 2017
% sun2017_computation, Oct. 30, 2017
% Kimcache2017, Oct. 30 2017 ~~~
% mirmohseni2017private, Nov. 13, 2017
% abdul2017private, Nov. 14, 2017
% wei2017jsac, Dec. 18 2017 ~~~
% xu2018building, Jan. 8 2018
% banawan2018asymmetry, Jan. 9 2018
% karpuk2018private, Jan. 13 2018
% banawan2018private, Jan. 18 2018
% chen2018asymptotic, Jan. 18 2018
% tajeddine2018private, March 2018
% wang2018secure, Apr. 26, 2018
% attia2018capacity, May 10, 2018
% wei2018capacity, Jun. 4, 2018
% tajeddine2018privatex, Jun. 20, 2018
% banawan2018noisy, Jul. 16, 2018
% jia2018cross, Aug. 22, 2018
% tian2018capacity, Aug. 24 2018
% kumar2018private, Sep. 4 2018
% bitar2018staircase, Spe. 4, 2018
% li2018converse, Sep. 26, 2018
% d2018one, Oct. 12, 2018
% tajeddine2018private, Oct. 21, 2018

Coded caching refers to the problem of placing files in users' local storage caches ahead of time properly and designing efficient delivery schemes at the time of specific user requests in such a way to minimize the traffic during the delivery phase. In the original setup \cite{maddah2014fundamental} (see Fig.~\ref{caching_PIR}), a server with $K$ files connects to $N$ users through an error-free shared link, where each user has a local memory which can store up to $M$ files. The system operates in two phases, a placement phase and a delivery phase. In the placement phase, the server places the files into each user's local memory. In the delivery phase, each user requests a file from the server, and the server aims to satisfy all the requests with the lowest traffic load. If the set of users in the two phases are identical, the server can arrange the content in each user's local memory in an optimized manner, which is called \textit{centralized coded caching}. Reference \cite{maddah2014fundamental} proposes a symmetric batch caching scheme, which is shown to be optimal for the case of centralized uncoded placement in \cite{yu2016exact}. If the set of users in the two phases varies, the server cannot arrange the files in user caches in a centralized manner. Instead, the server treats each user identically and independently which is called  \textit{decentralized coded caching} \cite{maddah2015decentralized}. Reference \cite{maddah2015decentralized} proposes a uniform and random caching scheme, which is shown to be optimal for the case of decentralized uncoded placement in \cite{yu2016exact}. Many interesting variants of coded caching problem have been investigated in \cite{ji2016fundamental, pedarsani2016online, ghasemi2017improved, shanmugam2016finite, sengupta2015fundamental, zhang2017fundamental, xu2017fundamental, tian2018caching, bidokhti2018noisy, yu2018characterizing, ibrahim2018coded, hassanzadeh2018rate, wan2016optimality, yang2018coded, zewail2018combination}.

\begin{figure}[t]
	\centering
	\epsfig{file=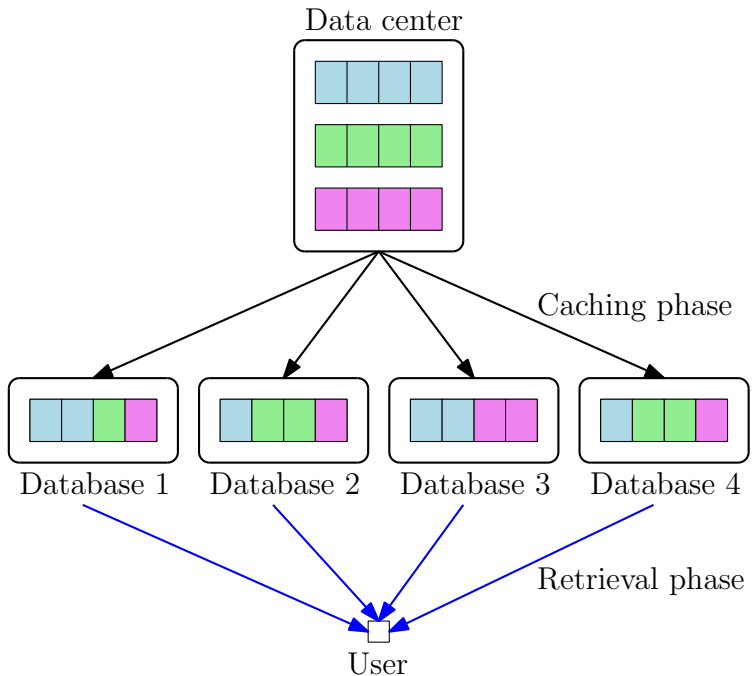,width=0.6\textwidth}
	\caption{PIR from centralized caching databases.}
	\label{central_PIR}
\end{figure}

The references that are most closely related to our work here are \cite{abdul2017private, attia2018capacity}. References \cite{abdul2017private, attia2018capacity} formulate a new type of PIR problem where the content of each database is not fixed a priori, but can be optimized to minimize the download cost. These papers bring PIR and coded caching problems together in a practically relevant and theoretically interesting manner. In their problem setting (see Fig.~\ref{central_PIR}), there is a data center (server) containing all the $K$ files where each file is of size $L$ bits, and the system operates in two phases. In the caching phase, there are $N$ databases in the system with a common storage size constraint $\mu$, i.e., each database can at most store $\mu K L$ bits, $ \frac{1}{N} \leq \mu \leq 1$. In the retrieval phase, a user (retriever) accesses the $N$ databases, and wishes to download a desired file privately. They consider the problem of optimally storing content from the data center to the databases in the caching phase in such a way that the normalized download cost during the retrieval phase is minimized. They focus on the \textit{centralized uncoded} caching case, i.e., the set of users in the two phases are identical so that the data center can assign the files to each database in a \textit{centralized} manner, and caching is \textit{uncoded} in that each database stores a subset of the bits from the data center (no coding), i.e., each database stores $\mu K L $ bits out of the total $KL$ bits. Surprisingly, they show that the symmetric batch caching scheme proposed in \cite{maddah2014fundamental} results in the lowest normalized download cost in the retrieval phase.

\begin{figure}[t]
	\centering
	\epsfig{file=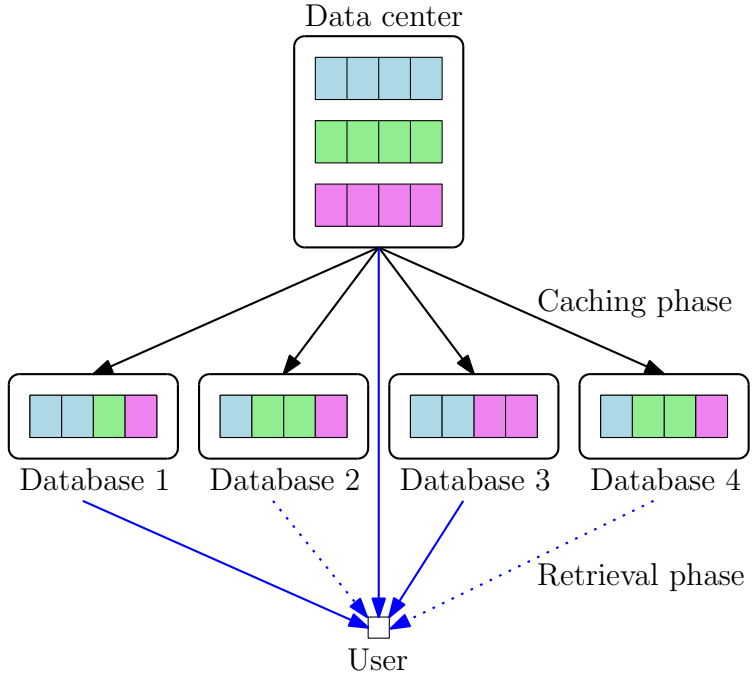,width=0.6\textwidth}
	\caption{PIR from decentralized caching databases.}
	\label{decentral_PIR}
\end{figure}

We consider the PIR problem from \textit{decentralized uncoded} caching databases. In our problem setting (see Fig.~\ref{decentral_PIR}), the system also operates in two phases as in \cite{abdul2017private, attia2018capacity}. However, the set of databases active in the two phases are different, and we do not know in advance which databases  the user (retriever) can access in the retrieval phase. Therefore, we consider a \textit{decentralized} setting for the caching phase, i.e., the data center treats each database identically and independently, or equivalently, each database chooses a subset of bits to store independently according to the same probability distribution. Here, we aim at designing the optimal probability distribution in the caching phase and PIR scheme in the retrieval phase such that the normalized download cost in the retrieval phase is minimized. Another main difference between our work and references \cite{abdul2017private, attia2018capacity} is that, in the caching phase, references \cite{abdul2017private, attia2018capacity} require that the $N$ databases altogether can reconstruct the entire $K$ files, i.e., when the user (retriever) connects to the $N$ databases, their collective content is equivalent to the content in the data center, so the user can download any desired file. While this can be guaranteed in the centralized setting, in the decentralized setting, where cache placement is probabilistic, we cannot guarantee that any given $N$ databases contain all the bits that exist in the data center. Thus, in order to formulate a meaningful PIR problem, we allow the user (retriever) access the data center as well as the databases in the retrieval phase. Finally, we remark about another sub-branch of PIR literature that considers caching: \cite{kadhe2017private, wei2017fundamental, chen2017capacity, wei2017capacity,  wei2017jsac, li2018converse}; there the user (retriever) itself has a cache memory where it stores a subset of the bits available in the databases. That problem is unrelated to the setting here even though it is also referred to as PIR with caching; in essence, it is PIR with side information.

In this work, for PIR from decentralized caching databases, we show that uniform and random caching scheme, originally proposed in \cite{maddah2015decentralized} for decentralized coded caching, results in the lowest expected normalized download cost in the retrieval phase. For the achievability, we apply the PIR scheme in \cite{JafarPIR} successively for all resulting subfile parts. For the converse, we first apply the lower bound derived in \cite{attia2018capacity}, which replaces the random variables for queries and answering strings by the content of the distributed databases in a novel manner extending the lower bounding techniques in \cite[Lemma~5 and Lemma~6]{JafarPIR}. To compare different probability distributions in the caching phase, we focus on the marginal distributions on each separate bit. Then, by using the nature of decentralization and uncoded caching, we further lower bound the normalized download cost. Finally, we show the matching converse for the expected normalized download cost to be $\frac{D}{L} =  \sum_{n=1}^{N+1}  \binom{N}{n-1} \mu^{n-1} (1-\mu)^{N+1-n} \left( 1+ \frac{1}{n} + \dots+ \frac{1}{n^{K-1}}  \right)$, which yields an exact capacity result for the problem.

\section{System Model}
We consider a system consisting of one data center and several databases. The data center stores $K$ independent files, labeled as $W_1$, $W_2$, $\dots$, $W_K$, where each file is of size $L$ bits. Therefore,
\begin{align}
H(W_1)=\dots=H(W_K)=L, \qquad H(W_1, \dots, W_K)=H(W_1)+\dots+H(W_K).
\end{align}
Each database has a storage capacity of $\mu K L$ bits, where $0\leq \mu \leq1$.

The system operates in two phases: In the caching phase, we consider the case of \textit{uncoded} caching, i.e., each database stores a subset of bits from the data center. Due to the storage size constraint, each database at most stores $\mu K L $ bits out of the total $KL$ bits from the data center. Here, we denote $i$th database as DB$_i$ and use random variable $Z_i$ to denote the stored content in DB$_i$. Therefore, the storage size constraint for DB$_i$ is
\begin{align} \label{eq_storage_size_constraint}
H(Z_i) \leq \mu K L .
\end{align}
We consider the \textit{decentralized} setting for the caching phase, i.e., each database chooses a subset of bits to store independently according to the same probability distribution, denoted by $P_H$. Rigorously, let random variable $H_i$ denote the indices of the stored bits in DB$_i$. For $N$ databases, the decentralized caching scheme $\mathcal{H}$ can be specified as
\begin{align} \label{eq_H}
\mathbb{P}(\mathcal{H}=(H_1, \dots, H_N))=\prod_{i=1}^N P_H(H_i).
\end{align}

In the retrieval phase, the user accesses $N$ databases and the data center. We note that we do not know in advance which $N$ databases are available or which $N$ databases the user will have access to. Here, we also assume that in the retrieval phase, the data center and $N$ databases do not communicate with each other (no collusion). To simplify the notation, we use DB$_0$ to denote the data center, and therefore $Z_0=(W_1, \dots, W_K)$ since the data center stores all the $K$ files. The user privately generates an index $\theta \in [K]=\{1, \dots, K\}$, and wishes to retrieve file $W_\theta$ such that it is impossible for either the data center or any individual database to identify $\theta$. For random variables $\theta$, and $W_1,\dots,W_K$, we have
\begin{align} \label{independency}
H\left(\theta, W_1,\dots,W_K  \right)= H\left( \theta \right)  + H(W_1)+\dots+H(W_K).
\end{align}
In order to retrieve file $W_\theta$, the user sends $N+1$ queries $Q_0^{[\theta]}, \dots, Q_{N}^{[\theta]}$ to DB$_0$, $\dots$, DB$_N$, where $Q_n^{[\theta]}$ is the query sent to DB$_n$ for file $W_\theta$. Note that the queries are independent of the realization of the $K$ files. Therefore,
\begin{align} \label{query_indep}
I(W_1, \dots, W_K;  Q_0^{[\theta]}, \dots,  Q_{N}^{[\theta]}  ) =0.
\end{align}
Upon receiving the query $Q_n^{[\theta]}$, DB$_n$ replies with an answering string $A_n^{[\theta]}$, which is a function of  $Q_n^{[\theta]}$ and $Z_n$. Therefore, $\forall \theta \in [K], \forall n \in \{0\} \cup [N]$,
\begin{align} \label{answer_constraint}
H(A_n^{[\theta]}|Q_n^{[\theta]}, Z_n)=0.
\end{align}

After receiving the answering strings $A_0^{[\theta]}, \dots, A_N^{[\theta]}$ from DB$_0$, $\dots$, DB$_N$, the user needs to decode the desired file $W_\theta$ reliably. By using Fano's inequality, we have the following reliability constraint
\begin{align} \label{reliability_constraint}
H\left(W_\theta|Q_0^{[\theta]}, \dots, Q_N^{[\theta]}, A_0^{[\theta]}, \dots, A_N^{[\theta]} \right) = o(L),
\end{align}
where $o(L)$ denotes a function such that $\frac{o(L)}{L} \rightarrow 0$ as $L \rightarrow \infty$.

To ensure that individual databases do not know which file is retrieved, we have the following privacy constraint, $\forall n \in \{0\} \cup [N]$, $\forall \theta \in [K]$,
\begin{align} \label{privacy_constraint}
(Q_n^{[1]}, A_n^{[1]}, W_1, \dots, W_K)  \sim (Q_n^{[\theta]}, A_n^{[\theta]}, W_1, \dots, W_K) ,
\end{align}
where $A \sim B$ means that $A$ and $B$ are identically distributed.

Given that each file is of size $L$ bits, for a fixed $K$, $\mu$ and decentralized caching probability distribution $P_H$, let $\mathcal{H}$ denote the indices of the cached bits in the $N$ databases available in the retrieval phase. The probability distribution of $\mathcal{H}$ is specified in \eqref{eq_H}. Let $D^{[\theta]}_{\mathcal{H}}$ represent the number of downloaded bits via the answering strings $A_{0:N}^{[\theta]}$, where $A_{0:N}^{[\theta]}=(A_0^{[\theta]}, \dots, A_N^{[\theta]} )$. Then,
\begin{align}
D_{\mathcal{H}}^{[\theta]}=\sum_{n=0}^N H\left(A_n^{[\theta]} \right).
\end{align}
We further denote $D_{\mathcal{H}}$ as the expected number of downloaded bits with respect to different file requests, i.e., $D_{\mathcal{H}}=E_\theta\left[D_{\mathcal{H}}^{[\theta]}\right]$. Finally, we denote $D$ as the expected number of downloaded bits with respect to different realization of the cached bit indices, i.e.,  $D=E_{\mathcal{H}} \left[D_{\mathcal{H}}\right]$. A pair $\left(D,L\right)$ is achievable if there exists a PIR scheme satisfying the reliability constraint \eqref{reliability_constraint} and the privacy constraint \eqref{privacy_constraint}. The optimal normalized download cost $D^*$ is defined as
\begin{align}
D^*=\inf \left\{ \frac{D}{L}: \left(D, L \right) \text{ is achievable}  \right\}.
\end{align}
In this work, we aim at characterizing the optimal normalized download cost and finding the optimal decentralized caching probability distribution. 

\begin{figure}[t]
	\centering
	\epsfig{file=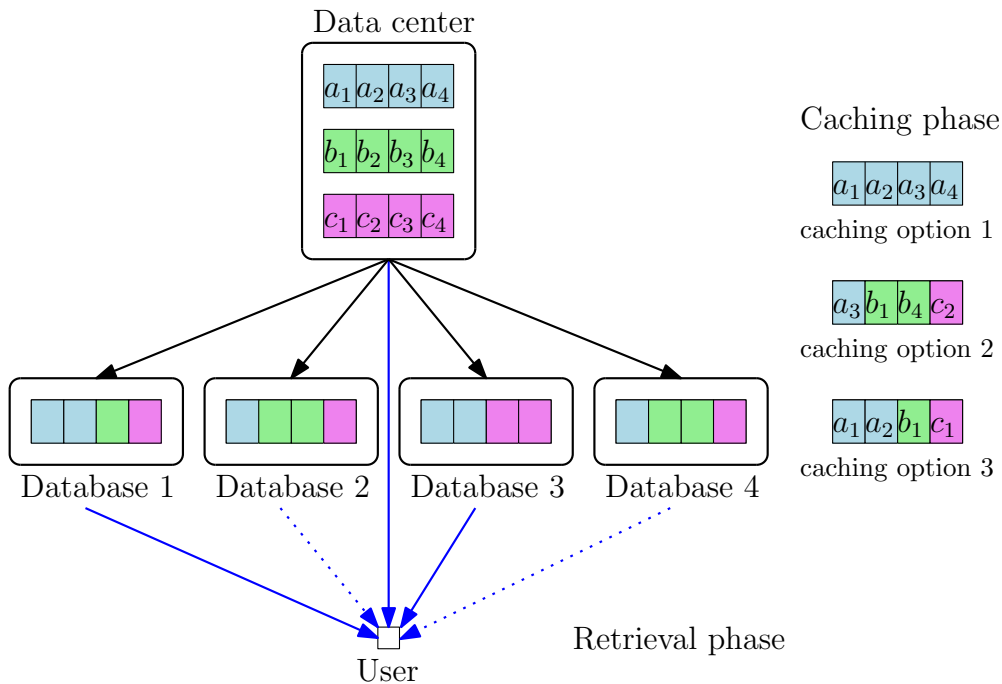,width=0.8\textwidth}
	\caption{PIR from decentralized caching databases with $K=3$, $N=2$, and $\mu=\frac{1}{3}$.}
	\label{ex_intro}
\end{figure}

Next, we illustrate the system model and the problem considered with a simple example of $K=3$ files and $N=2$ databases in the retrieval phase; see Fig.~\ref{ex_intro}. Consider a data center storing $K=3$ files where each file is of size $4$ bits. In the caching phase, there are $4$ databases in the system, and each database can at most store $4$ bits. Each database can always store the first file, which is of size $4$ bits, as caching option $1$ in  Fig.~\ref{ex_intro}. Or each database can uniformly and randomly choose $4$ bits out of total $12$ bits from the data center to store. One of the realization is shown as caching option $2$ in Fig.~\ref{ex_intro}. Each database can also choose $2$ bits from the first file and $1$ bit each from the remaining two files to store, where one of the realization is shown as caching option $3$ in Fig.~\ref{ex_intro}. We require each database to use the same probability distribution to choose the bits to store in order to satisfy the decentralized requirement. In this example, we assume that the user can access the data center and $N=2$ databases in the retrieval phase, say the first and the third database, and the user wishes to download a file privately. Our questions are as follows: What is the optimal probability distribution to use in the caching phase? What is the optimal PIR scheme to use in the retrieval phase? How can we jointly design the schemes in the two phases such that the expected normalized download cost is the lowest in the second phase?

\section{Main Results and Discussions}

We characterize the optimal normalized download cost for PIR from decentralized uncoded caching databases in the following theorem.

\begin{theorem}\label{thm1}
For PIR from decentralized uncoded caching databases with $K$ files, where each file is of size $L$ bits, $N$ databases in addition to a data center available in the retrieval phase, and a storage size constraint $\mu K L $, $0<\mu<1$, bits for each database, the optimal normalized download cost is
\begin{align} \label{eq_thm}
\frac{D}{L} =  \sum_{n=1}^{N+1}  \binom{N}{n-1} \mu^{n-1} (1-\mu)^{N+1-n} \left( 1+ \frac{1}{n} + \dots+ \frac{1}{n^{K-1}}  \right) .
\end{align}
\end{theorem}

The achievability scheme is provided in Section~\ref{Sec_Ach}, and the converse proof is shown in Section~\ref{Sec_conv}. We first use the following example to show the main ingredients of Theorem~\ref{thm1}.

\subsection{Motivating Example: $K=3$ and $N=2$}

In this example, we consider the case where the data center stores $K=3$ independent files labeled as $A$, $B$, and $C$, where each file is of size $L$ bits. In the caching phase, several databases with storage capacity of $3 \mu L$ bits are present in the system. We will show that the optimal normalized download cost is $\frac{D}{L}=\frac{17}{18}\mu^2-\frac{5}{2}\mu+3$ when $N=2$ databases in addition to the data center are available in the retrieval phase.

\subsubsection{Achievability Scheme}

In the caching phase, to satisfy the storage size constraint, each database randomly and uniformly stores $3 \mu L$ bits out of total $3L$ bits from the data center. Each database operates independently through the same probability distribution resulting in decentralized caching.

In the retrieval phase, suppose $N=2$ databases, labeled as DB$_1$ and DB$_2$, in addition to the data center, labeled as DB$_0$, are available to the user, and the user wishes to retrieve file $A$ privately. Let us first focus on one file, say $A$. We can partition file $A$ into four subfiles
\begin{align}
A=(A_0, A_{0,1}, A_{0,2}, A_{0,1,2}),
\end{align}
where, for $S \subseteq \{0,1,2\}$, $A_S$ denotes the bits of file $A$ which are stored in databases in $S$. For example, $A_0$ denotes the bits of file $A$ only stored in DB$_0$ and $A_{0,2}$ denotes the bits of file $A$ stored in DB$_0$ and DB$_2$ and so on. Since each bit is stored in the data center, $0$ exists in the label of every partition. By the law of large numbers,
\begin{align}
|A_S|=L\mu ^ {|S|-1}(1-\mu)^{3-|S|} + o(L),
\end{align}
when the file size is large enough. We can do the same partitions for files $B$ and $C$.

To retrieve file $A$ privately, we first retrieve the subfile $A_{0,1,2}$ privately. We apply the PIR scheme proposed in \cite{JafarPIR} to retrieve the subfile $A_{0,1,2}$. Subfile $A_{0,1,2}$ is replicated in $3$ databases and the total number of files is $3$ since we also have $B_{0,1,2}$ and $C_{0,1,2}$. Therefore, we download
\begin{align} \label{eq_012}
L \mu ^2 \left(1+\frac{1}{3}+\frac{1}{9}\right) + o(L)
\end{align}
bits. We also need to retrieve the subfile $A_{0,1}$ privately. Subfile $A_{0,1}$ is replicated in $2$ databases and the total number of files is $3$ since we also have $B_{0,1}$ and $C_{0,1}$. By applying the PIR scheme in \cite{JafarPIR}, we download
\begin{align} \label{eq_01}
L \mu (1-\mu) \left( 1+ \frac{1}{2} +\frac{1}{4}  \right) + o(L)
\end{align}
bits. Next, we need to retrieve the subfile $A_{0,2}$ privately. Using \cite{JafarPIR}, we download
\begin{align} \label{eq_02}
L \mu (1-\mu) \left( 1+ \frac{1}{2} +\frac{1}{4}  \right) + o(L)
\end{align}
bits. Finally, we need to retrieve $A_0$ privately. Using \cite{JafarPIR}, we download
\begin{align} \label{eq_0}
L (1-\mu)^2 (1+1+1) + o(L)
\end{align}
bits. By adding \eqref{eq_012}, \eqref{eq_01}, \eqref{eq_02} and \eqref{eq_0}, we show that the normalized download cost
\begin{align} \label{eq_ach}
\frac{17}{18}\mu^2-\frac{5}{2}\mu+3
\end{align}
is achievable.

\subsubsection{Converse Proof} 

Here, we show that among all the decentralized caching probability distributions $P_H$, the lowest normalized download cost for $N=2$ databases is as shown in \eqref{eq_ach}. Given a decentralized caching probability distribution $P_H$, we have a resulting $\mathcal{H}$ in the retrieval phase.

We lower bound $D_{\mathcal{H}}$ first. In the retrieval phase, the stored content of DB$_0$, DB$_1$, and DB$_2$ are fixed  and uncoded, i.e., $Z_0$, $Z_1$ and $Z_2$ are fixed  and uncoded. We can apply the lower bound in \cite[Eqn.~(31)]{attia2018capacity} as the lower bound for $D_{\mathcal{H}}$. Therefore,
\begin{align}
D_{\mathcal{H}} &\geq L + \frac{4}{27} \sum_{k=1}^3 H(W_k) + \frac{11}{108} \sum_{i=0}^2 \sum_{k=1}^3 H(W_k|Z_i) + \frac{17}{54} \sum_{i=0}^2 \sum_{k=1} ^ 3 H(W_k|Z_{[0:2] \setminus i}) +o(L) \\
&=\frac{13}{9}L + \frac{11}{108} \sum_{i=1}^2 \sum_{k=1}^3 H(W_k|Z_i) + \frac{17}{54} \sum_{k=1} ^ 3 H(W_k|Z_1, Z_2) +o(L) \label{eq_a} \\
&\geq \frac{13}{9}L + \frac{11}{108} \left( 3L-3 \mu L  + 3L-3 \mu L \right) + \frac{17}{54} \sum_{k=1} ^ 3 H(W_k|Z_1, Z_2) +o(L)\label{eq_b}  \\
&=\frac{37}{18}  L - \frac{11}{18}  \mu L + \frac{17}{54}  H(W_{1:3}|Z_1, Z_2) +o(L), \label{eq_c}
\end{align}
where \eqref{eq_a} holds due to $Z_0=(W_1,W_2,W_3)$, and \eqref{eq_b} holds due to \eqref{eq_storage_size_constraint}. We note that different $\mathcal{H}$ results in different $Z_1$ and $Z_2$.

We lower bound $D$ now. From \eqref{eq_c}, we have
\begin{align} \label{lb_1}
D=E_{\mathcal{H}}\left[D_{\mathcal{H}} \right] \geq \frac{37}{18}  L - \frac{11}{18}  \mu L + \frac{17}{54} E_{\mathcal{H}}\left[ H(W_{1:3}|Z_1, Z_2) \right] +o(L).
\end{align}
Let random variables $X_{i,j}^{(n)}$, $i=1, \dots, L$, $j=1, \dots, K$, be the indicator functions showing that the $i$th bit of file $W_j$ is cached in DB$_n$ or not, i.e., $X_{i,j}^{(n)}=1$ means that the $i$th bit of file $W_j$ is stored in DB$_n$ and $X_{i,j}^{(n)}=0$ means that it is not stored in DB$_n$. For DB$_1$ we have
\begin{align} \label{eq_xx}
X_{1,1}^{(1)}+\dots+X_{L,1}^{(1)} + X_{1,2}^{(1)}+\dots+X_{L,2}^{(1)} + X_{1,3}^{(1)}+\dots+X_{L,3}^{(1)} \leq 3 \mu L
\end{align}
due to the storage size constraint in \eqref{eq_storage_size_constraint}. We note that $P_H$ induces probability measures on random variables $X_{i,j}^{(n)}$, and let $X_{i,j}^{(n)}=1$ with probability $p_{i,j}$, where we remove the superscript $n$ since each database adopts the same probability distribution $P_H$ to choose the cached bits due to the decentralized property. By taking expectation on \eqref{eq_xx} and applying the linearity of expectation, we have
\begin{align}
E[X_{1,1}^{(1)}] + \dots + E[X_{L,3}^{(1)}] &\leq 3 \mu L, 
\end{align}
which yields
\begin{align}
p_{1,1}+\dots + p_{L,3} & \leq 3 \mu L.  \label{eq_yy}
\end{align}
Let random variables $V_{i,j}$, $i=1, \dots, L$, $j=1, \dots, K$, be the indicator functions showing that the $i$th bit of file $W_j$ is not cached in DB$_1$ and DB$_2$, i.e., $V_{i,j}=1$ means that the $i$th bit of file $W_j$ is not stored in either DB$_1$ or DB$_2$. Therefore, we have
\begin{align}
V_{i,j}=(1-X_{i,j}^{(1)})(1-X_{i,j}^{(2)}).
\end{align}
Now, we can evaluate $E_{\mathcal{H}}\left[ H(W_{1:3}|Z_1, Z_2) \right]$ in \eqref{lb_1} as follows
\begin{align}
E_{\mathcal{H}}\left[ H(W_{1:3}|Z_1, Z_2) \right] &= E[ V_{1,1}+\dots+V_{L,3}  ] \\
                                                  &= E[ V_{1,1}] + \dots + E[V_{L,3}] \\
                                                  &= (1-p_{1,1})^2 + \dots + (1-p_{L,3})^2.
\end{align}
Therefore, continuing from \eqref{lb_1}, we have
\begin{align} \label{lb_2}
D \geq  \frac{37}{18}  L - \frac{11}{18}  \mu L + \frac{17}{54}\left[  (1-p_{1,1})^2 + \dots + (1-p_{L,3})^2 \right] +o(L),
\end{align}
where $p_{1,1}$, $\dots$, $p_{L,3}$ are subject to \eqref{eq_yy}. To further lower bound the right hand side of \eqref{lb_2}, we minimize the right hand side with respect to $p_{i,j}$ subject to \eqref{eq_yy}. Hence, we consider the following Lagrangian
\begin{align}
L(p_{1,1},\dots,p_{L,3},\lambda)=(1-p_{1,1})^2 + \dots + (1-p_{L,3})^2 + \lambda \left( p_{1,1}+\dots + p_{L,3} - 3 \mu L    \right).
\end{align}
From the KKT conditions, we have
\begin{align}
\lambda = 2(1-p_{i,j}), \quad i=1, \dots, L, \quad j=1,2,3.
\end{align}
Thus, we can further lower bound \eqref{lb_2} by letting $p_{1,1}=\dots=p_{L,3}=\mu$, and we have
\begin{align}
\frac{D}{L} & \geq  \frac{37}{18}  - \frac{11}{18}  \mu  + \frac{17}{54}\left[ 3(1-\mu)^2  \right] + \frac{o(L)}{L}  \\
            & = \frac{17}{18}\mu^2-\frac{5}{2}\mu+3+\frac{o(L)}{L} .
\end{align}
Therefore, we show that the optimal normalized download cost is $\frac{17}{18}\mu^2-\frac{5}{2}\mu+3$ when $N=2$ databases in addition to the data center are available in the retrieval phase. To achieve the optimal normalized download cost, each database should randomly and uniformly store the bits in the caching phase.

\subsection{Further Examples and Numerical Results}

\begin{figure}
\centering
\epsfig{file=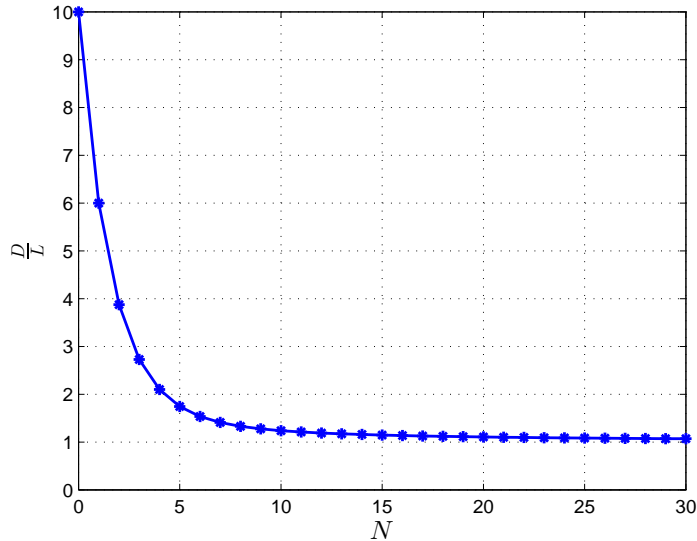,width=0.63\textwidth}
\caption{PIR from different number of available databases in the retrieval phase with $K=10$ and $\mu=\frac{1}{2}$.}
\label{K_10_diff_N}
\end{figure}

\begin{figure}
	\centering
	\epsfig{file=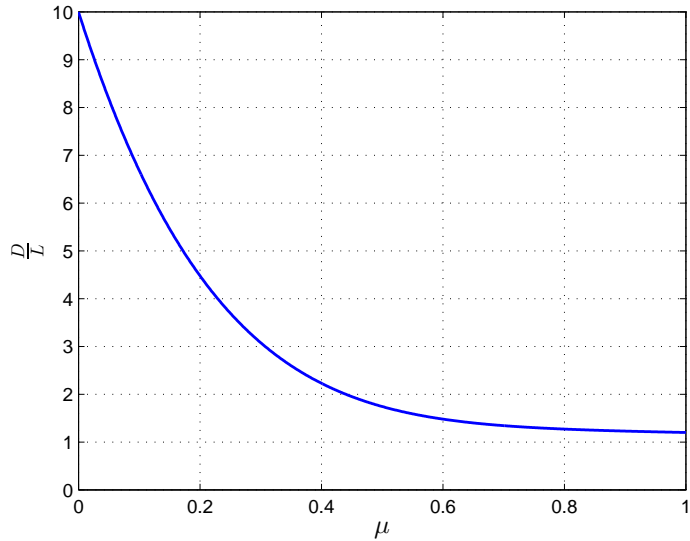,width=0.63\textwidth}
	\caption{PIR from $N=5$ databases with different storage constraint $\mu$ with $K=10$.}
	\label{K_10_N_5_diff_mu}
\end{figure}

Now, we use different scenarios to illustrate the optimal normalized download cost in \eqref{eq_thm}. We first consider the scenario where the data center contains $K=10$ files, each database with storage size constraint $\mu=\frac{1}{2}$, and in the retrieval phase, the user can access $N=0,\dots,30$ databases in addition to the data center. We plot the expected normalized download cost versus different number of available databases in Fig.~\ref{K_10_diff_N}. When $N=0$, in order to download the desired file privately, the user should download all the files in the data center, and this results in a download cost of $\frac{D}{L}=K=10$. As the number of accessible databases increases, the normalized download cost decreases. We next consider the scenario where the data center contains $K=10$ files, and the user can access $N=5$ databases in addition to the data center in the retrieval phase. We plot the expected normalized download cost versus different storage size constraint $\mu$ in Fig.~\ref{K_10_N_5_diff_mu}. When $\mu=0$, in order to download the desired file privately, the user should download all the files in the data center resulting in $\frac{D}{L}=K=10$. As $\mu$ increases, the normalized download cost decreases. Finally, we conclude this section with the following general remarks about our main result.

\subsection{Remarks}

\begin{remark}
The achievability scheme consists of two parts, the design of the probability distribution in the caching phase and the PIR scheme in the retrieval phase. We find that the uniform and random caching scheme, originally proposed in \cite{maddah2015decentralized} for decentralized coded caching, results in the optimal normalized download cost in the retrieval phase. We remark here that the symmetric batch caching scheme, originally proposed in \cite{maddah2014fundamental} for centralized coded caching, also results in the optimal normalized download cost for PIR from centralized uncoded caching databases \cite{attia2018capacity}. In the retrieval phase, according to the distribution of the subfiles, we apply the PIR scheme proposed in \cite{JafarPIR} for all subfiles to retrieve the desired file.
\end{remark}

\begin{remark}
For the converse, we first apply the lower bound derived in \cite{attia2018capacity} which introduces new ingredients in addition to the interference lower bound lemma and induction lemma in \cite[Lemma~5 and Lemma~6]{JafarPIR}. We note that in \cite{attia2018capacity} the authors replace random variables for queries and answering strings by the contents of the distributed databases in a novel way which is crucial for the converse. With this replacement, we can account for different cached content in the caching phase resulting in different lower bound in the normalized download cost in the retrieval phase. Due to the nature of uncoded caching, this replacement facilitates further lower bound. For the decentralized problem here, to compare different probability distributions in the caching phase, we focus on the marginal distributions on each bit. This transformation allows us to use linearity of expectation, and the nature of decentralization and uncoded caching to further lower bound the expected normalized download cost.
\end{remark}

\begin{figure}
	\centering
	\epsfig{file=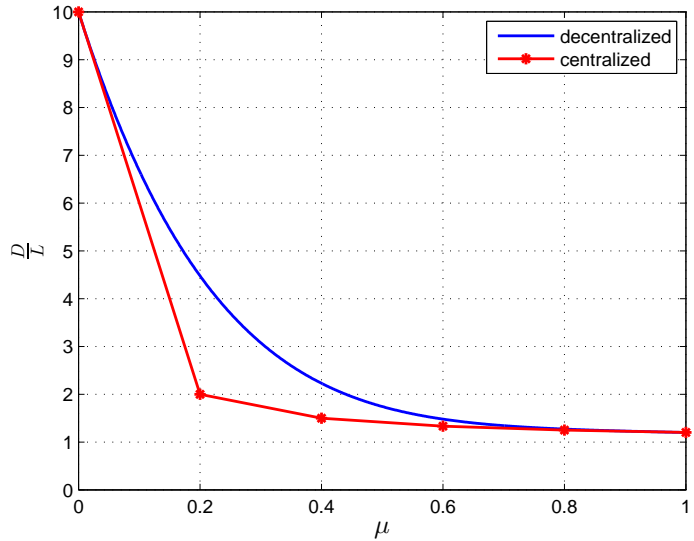,width=0.63\textwidth}
	\caption{PIR from centralized caching databases and decentralized caching databases.}
	\label{central_vs_decentral}
\end{figure}

\begin{remark}
A more directly related PIR problem from centralized uncoded caching databases for our setting is the one where, in the caching phase, the data center arranges the files in $N$ databases in a centralized manner, and in the retrieval phase, the user has access also to the data center in addition to the $N$ databases. This is different from the problem setting in \cite{abdul2017private, attia2018capacity}, since there the user can only access the $N$ databases in the retrieval phase. As a side note, we can show that symmetric batch caching scheme is still optimal for this extended problem setting where the data center also participates in the PIR stage. Rigorously, the optimal trade-off between storage and download cost in this case is given by the lower convex envelope of the following $(\mu, D(\mu))$ pairs, for $t=0, 1, \dots, N$,
\begin{align}
\left(\mu=\frac{t}{N}, D(\mu)=\sum_{k=0}^{K-1}\frac{1}{(t+1)^k}  \right).
\end{align}
To achieve this trade-off, the data center arranges the files into the $N$ databases as in \cite{abdul2017private, attia2018capacity}. In the retrieval phase, the user accesses also the data center; therefore, the subfiles are stored in one more database. For the converse, we no longer require all the $N$ databases to reconstruct the entire $K$ files as in \cite{abdul2017private, attia2018capacity}. Thus, while in \cite{abdul2017private, attia2018capacity} the smallest allowable $\mu$ is $\mu=\frac{1}{N}$, since the $N$ databases need to reconstruct the entire $K$ files, here since the user can access the data center, the parameter $\mu$ starts from $0$. Now, we can compare PIR from centralized caching databases and PIR from decentralized caching databases fairly, since in the retrieval phase, the user can access the data center in both cases. We consider the case where $K=10$ and $N=5$, and plot the result in Fig.~\ref{central_vs_decentral}.
\end{remark}

\section{Achievability Scheme} \label{Sec_Ach}

The achievability scheme consists of two parts: the design of the probability distribution used in the caching phase and the PIR scheme used in the retrieval phase. In the caching phase, each database uniformly and randomly stores $\mu K L$ bits from the data center. The storage size constraint in \eqref{eq_storage_size_constraint} is satisfied directly. Each database operates independently through the same probability distribution resulting in decentralized caching.

In the retrieval phase, suppose there are $N$ databases in addition to the data center available to the user. Each file $W_j$ can be expressed as
\begin{align}
W_j = \bigcup_{ \{0\} \subseteq S\subseteq \{0,1,\dots, N\}   }  W_{j,S},
\end{align}
where $W_{j,S}$ represents the bits of file $W_j$ which are stored in databases in $S$. Since each bit must be stored in the data center, i.e., DB$_0$, we have $\{0\} \subseteq S$. By the law of large numbers,
\begin{align}
|W_{j,S}| = L \mu^{|S|-1} (1-\mu)^{N+1-|S|}+o(L),
\end{align} 	
when the file size is large enough.

To retrieve the desired file, say $W_j$, privately, we retrieve each subfile, $W_{j,S}$, privately. Subfile $W_{j,S}$ is replicated in $|S|$ databases, and for each of these $|S|$ databases, there are $K$ subfiles, i.e., $W_{k,S}$, $k=1,\dots,K$. We apply the PIR scheme in \cite{JafarPIR} to retrieve $W_{j,S}$ privately by downloading
\begin{align}
L\mu^{|S|-1} (1-\mu)^{N+1-|S|} \left( 1+ \frac{1}{|S|} + \dots+ \frac{1}{|S|^{K-1}}  \right) +o(L)
\end{align}
bits. We also note that there are $\binom{N}{|S|-1}$ types of $W_{j,S}$. Therefore, the following normalized download cost
\begin{align} \label{eq_achf}
\frac{D}{L} =  \sum_{n=1}^{N+1}  \binom{N}{n-1} \mu^{n-1} (1-\mu)^{N+1-n} \left( 1+ \frac{1}{n} + \dots+ \frac{1}{n^{K-1}}  \right)
\end{align}
is achievable.

\section{Converse Proof} \label{Sec_conv}

We first derive a lower bound for $D_{\mathcal{H}}$. Since in the retrieval phase the content of DB$_0$, $\dots$, DB$_N$, are fixed to be $Z_0$, $\dots$, $Z_N$, we can use the lower bound derived in \cite[Eqn.~(71)]{attia2018capacity} to serve as the lower bound for $D_{\mathcal{H}}$. A key step to obtain \cite[Eqn.(71)]{attia2018capacity} is to replace the query and answering string random variables with the content of each database, i.e., replacement of $Q^{[k]}_\mathcal{N}$ and $A^{[k]}_\mathcal{N}$ with $Z_\mathcal{N}$. With this replacement, one can account for different cached content in the caching phase resulting in different lower bound in the normalized download cost in the retrieval phase. In addition, due to the nature of uncoded caching, this replacement facilitates a further lower bound. Moreover, to obtain \cite[Eqn.~(71)]{attia2018capacity}, the authors find interesting recursive relationships to compactly deal with the nested harmonic sums. Therefore, from \cite[Eqn.(71)]{attia2018capacity} we have
\begin{align} \label{eq_c_1}
D_{\mathcal{H}} \geq L + \sum_{l=1}^{N+1} \binom{N+1}{l} \left(  \frac{1}{l} + \frac{1}{l^2} + \dots+ \frac{1}{l^{K-1}} \right) x_l,
\end{align}
where
\begin{align} \label{eq_x_l}
x_l \triangleq \frac{1}{K \binom{N+1}{l}}  \sum_{  \{0\} \subseteq S \subseteq [0:N],~|S|=l} H(W_{1:K,S}), \quad l \in [1:N+1],
\end{align}
and $W_{1:K,S}$ represents the bits of files $W_{1:K}$ which are stored in databases in $S$. 

In the following lemma, we develop a lower bound for $E[x_l]$.
\begin{lemma}  \label{lemma_1}
For $l \in [1:N+1]$, and $x_l$ given in \eqref{eq_x_l}, we have 	
\begin{align}
E[x_l] \geq L   \mu^{l-1}(1-\mu)^{N+1-l}  \frac{\binom{N}{l-1}}{\binom{N+1}{l}}.
\end{align}	
\end{lemma}	
\begin{Proof}
By taking expectation on \eqref{eq_x_l} and using the linearity of expectation, we have
\begin{align} \label{lb_exp_1}
E[x_l] = \frac{1}{K \binom{N+1}{l}} \sum_{\{0\} \subseteq S \subseteq [0:N],~|S|=l} E[H(W_{1:K,S})].
\end{align}
Let random variables $X_{i,j}^{(n)}$, $i=1, \dots, L$, $j=1, \dots, K$, be the indicator functions showing that the $i$th bit of file $W_j$ is cached in DB$_n$, $n=0, \dots, N$, or not, i.e., $X_{i,j}^{(n)}=1$ means that the $i$th bit of file $W_j$ is stored in DB$_n$ and $X_{i,j}^{(n)}=0$ means that it is not stored in DB$_n$. For DB$_n$ we have
\begin{align} \label{eq_xx_n}
X_{1,1}^{(n)}+\dots+X_{L,1}^{(n)} + \dots + X_{1,K}^{(n)}+\dots+X_{L,K}^{(n)} \leq  \mu K L
\end{align}
due to the storage size constraint in \eqref{eq_storage_size_constraint}. We note that $P_H$ induces probability measures on random variables $X_{i,j}^{(n)}$, and let $X_{i,j}^{(n)}=1$ with probability $p_{i,j}$, where we remove the superscript $n$ since each database adopts the same probability distribution $P_H$ to choose the cached bits due to the decentralized caching property. By taking expectation on \eqref{eq_xx_n} and applying the linearity of expectation, we have
\begin{align}
E[X_{1,1}^{(n)}] + \dots + E[X_{L,K}^{(n)}] &\leq  \mu K L,
\end{align}
which yields
\begin{align}
p_{1,1}+\dots + p_{L,K} & \leq  \mu K L.  \label{eq_yy_n}
\end{align}
Let random variables $Y_{i,j}^S$, $i=1, \dots, L$, $j=1, \dots, K$, be the indicator functions showing that the $i$th bit of file $W_j$ is cached in DB$_n$, $n \in S$, i.e., $Y_{i,j}=1$ means that the $i$th bit of the file $W_j$ is stored in DB$_n$, $n \in S$. Therefore, we have
\begin{align}
Y_{i,j}^S= \prod_{n\in S} X_{i,j}^{(n)} \prod_{n\in [0:N] \setminus S}(1- X_{i,j}^{(n)}) .
\end{align}
Now, we can evaluate $E\left[ H(W_{1:K,S})  \right]$ in \eqref{lb_exp_1} as follows
\begin{align}
E\left[ H(W_{1:K,S})  \right] &= E[ Y_{1,1}^S+\dots+Y_{L,K}^S  ] \\
&= E[ Y_{1,1}^S] + \dots + E[Y_{L,K}^S] \\
&= p_{1,1}^{|S|-1}(1-p_{1,1})^{N+1-|S|} + \dots + p_{L,K}^{|S|-1}(1-p_{L,K})^{N+1-|S|} ,
\end{align}
where $p_{1,1}$, $\dots$, $p_{L,K}$ are subject to \eqref{eq_yy_n}. Now, continuing from \eqref{lb_exp_1}, we have
\begin{align}
E[x_l]& = \frac{1}{K \binom{N+1}{l}} \sum_{ \{0\} \subseteq  S \subseteq [0:N],~|S|=l}p_{1,1}^{l-1}(1-p_{1,1})^{N+1-l} + \dots + p_{L,K}^{l-1}(1-p_{L,K})^{N+1-l}    . \label{tmp2}
\end{align}
To further lower bound \eqref{tmp2}, we consider the following Lagrangian
\begin{align}
L(p_{1,1},\dots,p_{L,K},\lambda)&=p_{1,1}^{l-1}(1-p_{1,1})^{N+1-l} + \dots + p_{L,K}^{l-1}(1-p_{L,K})^{N+1-l} \notag \\
&\quad+ \lambda \left( p_{1,1}+\dots + p_{L,K} -  \mu K L    \right).
\end{align}
From the KKT conditions, we have
\begin{align}
\lambda = p_{i,j}^{l-1}(N+1-l)(1-p_{i,j})^{N-l}-(l-1)p_{i,j}^{l-2}(1-p_{i,j})^{N+1-l},
\end{align}
where $i=1, \dots, L$, $j=1,\dots,K$. Therefore, we can further lower bound \eqref{tmp2} by letting $p_{1,1}=\dots=p_{L,K}=\mu$, then we have
\begin{align}
E[x_l] &\geq \frac{1}{K\binom{N+1}{l}} \sum_{\{0\} \subseteq  S \subseteq [0:N],~|S|=l}  KL   \mu^{l-1}(1-\mu)^{N+1-l}  \\
       &=L   \mu^{l-1}(1-\mu)^{N+1-l}  \frac{\binom{N}{l-1}}{\binom{N+1}{l}},
\end{align}
which completes the proof.
\end{Proof}

Finally, by taking expectation and applying Lemma~\ref{lemma_1} to \eqref{eq_c_1}, we obtain
\begin{align}
\frac{D}{L} &\geq 1 + \sum_{l=1}^{N+1} \binom{N}{l-1} \left(  \frac{1}{l} + \frac{1}{l^2} + \dots+ \frac{1}{l^{K-1}}  \right)\mu^{l-1}(1-\mu)^{N+1-l} \\
& = \left(\mu+(1-\mu)\right)^N + \sum_{l=1}^{N+1} \binom{N}{l-1} \left(  \frac{1}{l} + \frac{1}{l^2} + \dots+ \frac{1}{l^{K-1}}  \right)\mu^{l-1}(1-\mu)^{N+1-l} \\
& =  \sum_{l=1}^{N+1} \binom{N}{l-1} \left( 1+ \frac{1}{l} + \frac{1}{l^2} + \dots+ \frac{1}{l^{K-1}}  \right)\mu^{l-1}(1-\mu)^{N+1-l}
\end{align}
which matches \eqref{eq_achf}.

\section{Conclusion}

We considered the PIR problem from decentralized uncoded caching databases. Due to the nature of decentralization and the storage size constraint, we allow the user to access the data center in the retrieval phase to guarantee that the user can reconstruct the entire desired file. We showed that uniform and random decentralized caching scheme, originally proposed in \cite{maddah2015decentralized} for the problem of decentralized coded caching, results in the lowest expected normalized download cost in the PIR phase. We characterized the expected normalized download cost to be $\frac{D}{L} =  \sum_{n=1}^{N+1}  \binom{N}{n-1} \mu^{n-1} (1-\mu)^{N+1-n} \left( 1+ \frac{1}{n} + \dots+ \frac{1}{n^{K-1}}  \right) $. For the achievability, we applied the PIR scheme in \cite{JafarPIR} for all subfiles. For the converse, we first applied the lower bound derived in \cite{attia2018capacity}, and to compare different probability distributions in the caching phase, we focused on the marginal distributions on individual bits. By using the nature of decentralization and uncoded caching, we further lower bounded the normalized download cost. Finally, we showed the matching converse for the expected normalized download cost, obtaining the exact capacity of the resulting PIR problem.

\bibliographystyle{unsrt}
\bibliography{references}
\end{document}